\documentclass[aps,prl,twocolumn,nofootinbib,groupedaddress,amsfonts,floatfix]{revtex4}

\usepackage{graphicx,amsmath,amssymb,amstext}
\usepackage{amssymb,amsbsy,amsfonts,amsthm,hyperref}
\usepackage{lscape,placeins}
\usepackage{url}

\newcommand{\adsurl}[1]{\href{#1}{ADS}}
\providecommand{\url}[1]{\href{#1}{#1}}

\newcommand{\be}{\begin{equation}}
\newcommand{\ee}{\end{equation}}
\newcommand{\bea}{\begin{eqnarray}}
\newcommand{\eea}{\end{eqnarray}}
\newcommand{\nn}{\nonumber}

\usepackage{color}
\usepackage{ifthen}
\newboolean{editorial}
\setboolean{editorial}{true}
\newcommand{\editorial}[2]{\ifthenelse{\boolean{editorial}}{\textcolor{red}{[\textsf{\textbf{{#1}}}: }\textcolor{blue}{\textsf{{#2}}}\textcolor{red}{]}}{}}

\begin{document}

\title{The Linear Point: A cleaner cosmological standard ruler}

\author{Stefano Anselmi ${}^{1,2,3}$}
\email[]{stefano.anselmi@iap.fr}
\author{Glenn D.~Starkman${}^{1}$}
\author{Pier-Stefano Corasaniti${}^{2}$}
\author{Ravi K.~Sheth${}^{4,5}$}
\author{Idit Zehavi${}^{6}$}

\affiliation{${}^1$Department of Physics/CERCA/Institute for the Science of Origins, Case Western Reserve University, Cleveland, OH 44106-7079 -- USA}
\affiliation{${}^2$ LUTH, UMR 8102 CNRS, Observatoire de Paris, PSL Research University, Universit\'e Paris Diderot, 92190 Meudon -- France}
\affiliation{${}^3$ Institut d'Astrophysique de Paris, CNRS UMR 7095 and UPMC, 98bis, bd Arago, F-75014 Paris -- France}
\affiliation{${}^4$ Center for Particle Cosmology, University of Pennsylvania, 209 S. 33rd St., Philadelphia, PA 19104 -- USA}
\affiliation{${}^5$ The Abdus Salam International Center for Theoretical Physics, Strada Costiera, 11, Trieste 34151 -- Italy}
\affiliation{${}^6$ Department of Astronomy, Case Western Reserve University, Cleveland, OH 44106-7079 -- USA}

\begin{abstract}
We show how a characteristic length scale imprinted in the galaxy two-point correlation function, dubbed the {\it linear point}, can serve as a comoving cosmological standard ruler. In contrast to the Baryon Acoustic Oscillation peak location, this scale is constant in redshift and is unaffected by non-linear effects to within $0.5$ percent precision. We measure the location of the linear point in the galaxy correlation function of the LOWZ and CMASS samples from the Twelfth Data Release (DR12) of the Baryon Oscillation Spectroscopic Survey (BOSS) collaboration. We combine our linear-point measurement with cosmic-microwave-background constraints from the Planck satellite to estimate the isotropic-volume distance $D_{V}(z)$, without relying on a model-template or {\it reconstruction} method. We find $D_V(0.32)=1264\pm 28$ Mpc and $D_V(0.57)=2056\pm 22$ Mpc respectively, consistent with the quoted values from the BOSS collaboration. This remarkable result suggests that all the distance information contained in the baryon acoustic oscillations can be conveniently compressed into the single length associated with the linear point.
\end{abstract}

\maketitle

There is widespread consensus that Baryon Acoustic Oscillations (BAO) in the large-scale distribution of galaxies can be used to infer cosmic distances, thus providing insight on the nature of dark energy in the universe (see e.g., \cite{
Bassett:2009mm, 2005ApJ...633..560E}). Future galaxy surveys such as Euclid,\footnote{\url{http://sci.esa.int/euclid/}} DESI,\footnote{\url{http://desi.lbl.gov}} and WFIRST\footnote{\url{https://wfirst.gsfc.nasa.gov}} have been designed to measure the BAO signal in galaxy-clustering observables with statistical errors of a few percent. However, at this level of precision the BAO imprint differs from the linear-theory prediction due to non-linear growth of the late-time clustering of the matter (see e.g., \cite{Crocce:2007dt, 2008PhRvD..77d3525S, 2014MNRAS.440.1420R, Anselmi:2012cn}). 

Non-linear effects smear out the amplitude of the BAO signal and shift the location of the BAO peak in the galaxy two-point correlation function, or equivalently, they damp and modify the locations of the BAO oscillations in the galaxy power spectrum. These effects can introduce systematic errors in the estimation of cosmic distances, and consequently in inferred cosmological parameters. A $1\%$ error in peak position leads to a $4\%$ error in the determination of the dark-energy equation of state at redshift $z=1$ (see e.g., \cite{Angulo:2007fw}).  Because of this, a number of techniques have been developed to standardize BAO distance measurements, although at the expense of introducing numerous caveats. Recent work by some of us has shown that there exists a characteristic point in the two-point correlation function on BAO scales, which we dubbed the {\it linear point} (LP), that it is largely insensitive to non-linear effects \cite{2016MNRAS.455.2474A}.

In this letter, we demonstrate that the LP can  be used cleanly and simply as a cosmological comoving standard ruler without reconstruction or other theory-rich post-processing of the observational data. Thus, the LP restores the BAO to its originally envisaged status as a standard ruler, rather than a standardizable one.

Early in the history of the universe, overdensities in the nearly homogeneous dark matter, by then decoupled from the plasma of ordinary baryonic matter and radiation, began to collapse under gravity. This collapse generated spherical acoustic waves, which propagated outward from the collapsing overdensities. As the universe cooled, the photons in the plasma eventually decoupled from the baryons, and diffused away from the concentrations of baryons and dark matter. Meanwhile the baryons' momentum redshifted away leaving them nearly in place in the final location of the acoustic wavefront. The result was baryon and dark matter overdensities at the  locations both of the original inhomogeneities and of the spherical wavefronts of the outward-going acoustic waves. These overdensities became preferred sites for galaxy formation. We are able therefore to observe the relic traces of these acoustic waves both in the cosmic microwave background (CMB) -- the photons that decoupled from the plasma -- and in the spatial correlations of cosmic structures, which were assembled much later.

Because the overdensities all began to collapse at essentially the same time, and because the acoustic waves were propagating through a homogeneous background, the size that the spherical wavefront reached, known as the sound horizon, is universal -- the same for each sourcing overdensity.  It depends, to be sure, on the expansion history of the universe during that epoch, and on the properties of the cosmological plasma, but not on the spectrum or amplitude of the fluctuations.

The initial overdensities and the resulting acoustic waves were of low amplitude, so the relevant early-universe physics is very nearly linear. The properties of the acoustic oscillations and their impact on the CMB have therefore been computed precisely and accurately as a function of cosmological parameters. Measurements of the CMB temperature and polarization power spectra have consequently afforded us quite precise estimates of those cosmological parameters.

The evolution of cosmic structure is more complicated. Galaxies, the leading tracers of the BAO, are non-linear structures. Determining the precise relation between the measured galaxy correlation function or power spectrum and the primordial power spectrum and other cosmological properties therefore involves understanding the non-linear growth of dark matter and associated baryonic structures, with all of the attendant mode-coupling and complicated baryonic physics.

A variety of techniques have been developed to extract information from the BAO (see e.g., \cite{2008ApJ...686...13S, 2007ApJ...664..675E, 2009MNRAS.400.1643S, 2012MNRAS.427.2132P}). However, to achieve the accuracy and precision necessary to make the BAO constraints relevant to modern cosmology, all current methods involve modeling the effects of non-linear physics on the BAO \cite{2008PhRvD..77d3525S, 2007PhRvD..75f3512S}. For example, reconstruction methods \cite{2007ApJ...664..675E,2012MNRAS.427.2132P} correct for non-linear effects by displacing galaxies using the Zeldovich approximation. While this enhances the signal-to-noise, and so leads to a more precise determination of cosmic distances, it comes at the cost of introducing model dependencies. Even when validated with the aid of N-body simulation analyses, these do not correctly propagate the variations of non-linear effects with cosmological parameters.

Alternatively, one may search for features at BAO scales that are insensitive to the non-linearities of the matter distribution. The LP introduced in \cite{2016MNRAS.455.2474A} is such a feature; it is well-approximated by the mid-point between the position of the peak and the dip in the two-point correlation function of the matter-density field as well as of biased tracers such as dark matter halos. In particular, in N-body simulations of a $\Lambda$CDM model, the amplitude of the correlation function at the LP is only very weakly affected by non-linearities and scale-dependent bias, while its comoving position remains unaltered compared to the linear-theory prediction at percent-level precision, independent of the normalisation amplitude of the primordial power spectrum and the scalar spectral index. This is why the LP can serve as a clean standard ruler.  In the following, we show how the LP position can be estimated from galaxy-clustering data in a manner that is free of theoretical model assumptions.

Let us consider the monopole term $\xi_0$ of the two-point correlation function of galaxies. This is usually expressed in terms of the comoving separation $s(z)$. However, comoving distances are not directly measured; rather, galaxy redshifts are converted into comoving space (real space) by assuming a fiducial cosmology. Then, the data are fit to a fixed fiducial template $\xi_0^{\rm fixed}(\alpha\,s_{\rm fid}(z))$ to determine the shift parameter $\alpha$ from which cosmic distance constraints are inferred (see e.g., \cite{2008ApJ...686...13S, 2012MNRAS.427.2146X}).  

The dependence on the fiducial cosmology can be avoided altogether by working with the rescaled variable $y\equiv s(z)/D_V(z)$, where $D_V(z)$ is the conventional isotropic-volume distance estimator 
\be
D_{V}(z) \equiv \left[(1+z)^{2}D_{A}(z)^{2}\frac{cz}{H(z)}\right]^{{1/3}}\,,
\label{dv}
\ee
where $H(z)$ is the Hubble rate, $D_A(z)$ is the angular-diameter distance, and $c$ is the speed of light. As pointed out in \cite{Sanchez:2012sg}, in this rescaled variable the relation
\be
\label{eqn:cosmologyindependent}
\xi_0 (s^{\rm fid}(z)/D_V^{\rm fid}(z)) \simeq \xi_0 (s^{\rm true}(z)/D_V^{\rm true}(z))
\ee
holds to a very good approximation. Corrections due to the Alcock-Paczynski effect are negligible on BAO scales, provided that the fiducial value of the cosmic matter density is sufficiently close to the true value \cite{2013MNRAS.431.2834X}. It is worth remarking that Eq.~(\ref{eqn:cosmologyindependent}) is a purely geometric statement that applies only when $s^{\rm fid}(z)$ and $s^{\rm true}(z)$ are extracted from data without reference to a cosmology-dependent theoretical template $\xi_0^{\rm th}$. 

\begin{figure}
\centering
\includegraphics[width=1\hsize]{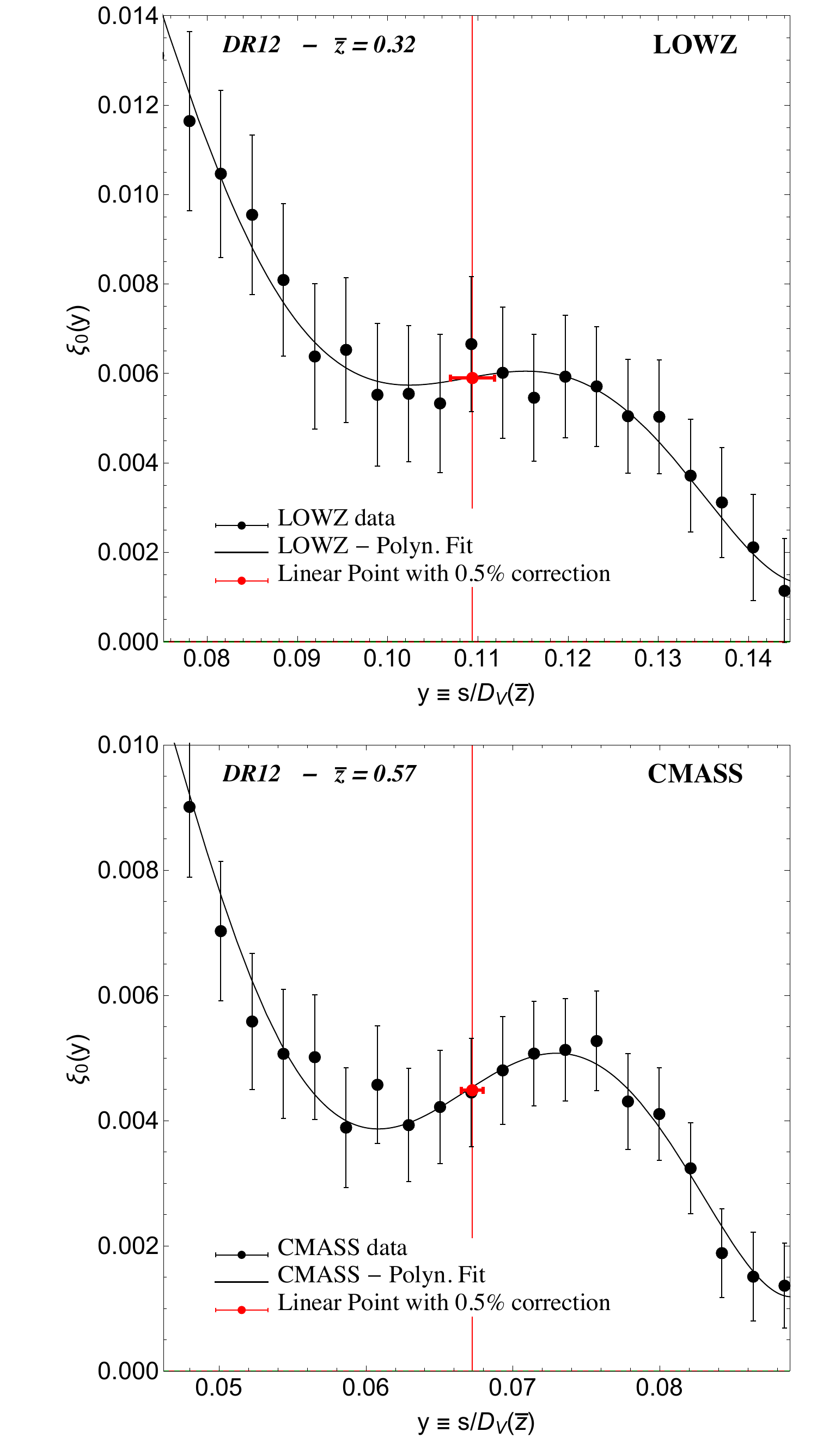}
\caption{
\label{fig:DR12} 
Linear point measurement in the galaxy clustering correlation function monopole. Black dots  show the LOWZ and CMASS correlation functions released by the BOSS collaboration \cite{2016MNRAS.457.1770C}. Red dots  represent the linear point measurements and the continuous black line the quintic polynomial $\xi_{0}(y)$ interpolation needed to extract it.}
\end{figure}

Since the LP identifies a comoving scale in the correlation function that does not evolve (i.e., it is constant in comoving coordinates) and that is immune to non-linear effects, we can exploit Eq.~(\ref{eqn:cosmologyindependent}) in combination with CMB constraints to infer the cosmic distance at the central redshift  $\bar{z}$ of the galaxy-survey sample. More precisely, we can estimate $s^{\rm true}_{LP}(\bar{z})$ using CMB information, since $s^{\rm true}_{LP}(\bar{z})=s^{\rm true}_{LP}(z_{\rm lin})$, where $s^{\rm true}_{LP}(z_{\rm lin})$ is the linear point in the correlation function predicted from the linear theory (say for the best-fit Planck cosmological model\footnote{In practice we compute $s^{\rm true}_{LP}(z_{\rm lin})$ from the linear correlation function at sufficiently high redshift where the linear theory is valid (e.g., $z_{\rm lin}=10$).}). Substituting in Eq.~(\ref{eqn:cosmologyindependent}),
\be
\label{standardruler}
\frac{s^{\rm true}_{LP}(z_{\rm lin})}{D_V^{\rm true}(\bar{z})}=\frac{s^{\rm fid}_{LP}(\bar{z})}{D_V^{\rm fid}(\bar{z})}= y_{LP}(\bar{z})\,,
\ee
where $y_{LP}(\bar{z})$ is the location of the rescaled LP measured in the galaxy correlation function at $\bar{z}$. Finally, the above equation can be inverted to infer the value of $D_V^{\rm true}(\bar{z})$. 

We remark that $s_{LP}^{\rm true}(z_{\rm lin})$ is insensitive to the late-time acceleration of the Universe. On the other hand, $y_{\rm LP} (\bar{z})$ strongly depends on $D_{V}$, and hence on the acceleration at redshift $\bar{z}$. This is the relevant feature of a comoving standard ruler.

In the following, we use the LP to estimate the isotropic-volume distance from the two-point correlation functions of the LOWZ and CMASS galaxy samples obtained by the BOSS collaboration.\footnote{\url{https://www.sdss3.org/surveys/boss.php}} 
In particular, we use the ``pre-reconstruction'' Data Release 12 (DR12) monopole correlation function presented in \cite{2016MNRAS.457.1770C}.\footnote{In the final galaxy clustering analysis performed by the BOSS collaboration  \cite{2017MNRAS.470.2617A} the galaxy sample survey volumes were chosen to be equal. Here we focus on the CF analysis presented in \cite{2016MNRAS.457.1770C}, which allows us to test the LP estimation procedure for different survey volumes as provided by the LOWZ and CMASS samples.}

Galaxy redshifts and angles have been converted into comoving distances assuming a $\Lambda$CDM fiducial cosmology specified by the following parameter values: $\Omega_{m}=0.29$, $\Omega_{b}h^{2}=0.02247$, $h=0.7$, $n_{s}=0.97$ and $\sigma_{8}=0.8$. 
The observed galaxy positions have not been corrected for peculiar motions. 
The central redshift of the LOWZ and CMASS samples are respectively $\bar{z}=0.32$ and $0.57$.

We convert comoving distances $s$ to $y$ using the BOSS fiducial cosmology. The correlation-function measurements around the BAO scales are shown in Fig.~\ref{fig:DR12}. We fit each $\xi_{0}(y)$ with a quintic polynomial and define the linear point as the midpoint of the peak and the dip in $\xi_{0}(y)$. We refer the reader to a companion paper \cite{LP_prd} for a detailed description of our model-independent estimation of the LP, which we validated using synthetic data. As discussed in \cite{2016MNRAS.455.2474A}, we increase this value of LP by $0.5\%$ to allow for a small secular evolution of $s_{LP}$ from high to low redshifts. This caps the systematic error on the determination of the LP to $0.5\%$ over the full range of redshifts. We find $y_{LP}({\bar z}_{\rm LOWZ-DR12}=0.32)=0.1094 \pm 0.0024$ and $y_{LP}({\bar z}_{\rm CMASS-DR12}=0.57)= 0.06724 \pm 0.00073$.

We estimate $s^{\rm true}_{LP}(z_{\rm lin})$ in Eq.~(\ref{standardruler}) for a $\Lambda$CDM cosmology best-fit to the Planck-TT,TE,EE+lowP anisotropy power spectra \cite{2016A&A...594A..13P} from the linear matter correlation function $\xi_{0}(s)$ computed using the CAMB code \cite{2000ApJ...538..473L}. We find $s_{LP}^{\rm Planck}=138.24$ Mpc. We neglect statistical errors due to the propagation of the Planck cosmological parameter uncertainties, which are expected to be negligible within the $\Lambda$CDM scenario (as in the case of the sound-horizon scale $r_{d}$).\footnote{Notice that both $s_{LP}$ and $r_{d}$ are independent of the values of the parameters describing the power spectrum of the primordial density fluctuations \cite{2016MNRAS.455.2474A} in inflationary $\Lambda$CDM, so the errors in $r_{d}$ and $s_{LP}$ are expected to be of the same order.} Then, from Eq.~(\ref{standardruler}),  $D_V^{\rm LP}(\bar{z})=s_{LP}^{\rm Planck}/y_{LP}({\bar z})$ and we derive 
\begin{eqnarray}
D_V^{\rm LP}(\bar{z}_{\rm LOWZ}=0.32) &=& (1264 \pm 28)\, {\rm Mpc}\, \nn \\
D_V^{\rm LP}(\bar{z}_{\rm CMASS}=0.57) &=& (2056 \pm 22)\, {\rm Mpc}\, . \nn
\end{eqnarray}

It is worth comparing these results to those obtained by the BOSS collaboration using the standard BAO method \cite{2016MNRAS.457.1770C}. For the same pre-reconstruction data as employed here, 
they quote\footnote{These values are obtained assuming $(r_d/r_d^{{\rm fid}})=1.00136$ with $r_d$ given by the Planck-TT,TE,EE+lowP best-fit values of cosmological parameters and $r_d^{{\rm fid}}=147.1$ Mpc in the \cite{2016MNRAS.457.1770C} fiducial cosmology.}
\bea
D_V^{\rm BOSS;pre-recon}(\bar{z}_{\rm LOWZ}=0.32) &=& (1247 \pm 37) {\rm Mpc}  \, \nn \\
D_V^{\rm BOSS;pre-recon}(\bar{z}_{\rm CMASS}=0.57) &=& (2043 \pm 27) {\rm Mpc}  .\,\nn
\eea
Hence the LP provides distance estimates with statistical uncertainties that are $24\%$ and $18\%$ smaller than the standard approach for the LOWZ and CMASS samples respectively. In the case of the ``post-reconstruction'' data, the BOSS collaboration report \cite{2016MNRAS.457.1770C}
\bea
D_V^{\rm BOSS;post-recon}(\bar{z}_{\rm LOWZ}=0.32) &=& (1265 \pm 21) {\rm Mpc}  \, \nn \\
D_V^{\rm BOSS;post-recon}(\bar{z}_{\rm CMASS}=0.57) &=& (2031 \pm 20) {\rm Mpc}\, ,\nn
\eea
thus with smaller statistical uncertainties than those found with the LP. However this is at the price of relying on the information contained in model assumptions built into the reconstruction method to synthetically increase the BAO signal-to-noise ratio.

Our results clearly demonstrate that cosmic distances can be inferred without relying heavily on a fiducial cosmology, using model-dependent templates, or having to manipulate the data so as to remove cosmology-dependent non-linear effects. Moreover, when the same pre-reconstruction data are fit, the LP provides remarkably smaller statistical errors than those reported by the BOSS collaboration. 

In a parallel study, employing the mock catalogues developed by the BOSS collaboration to reproduce the DR12 clustering properties, we validate the model-independent procedure exploited here to estimate the LP position from the observed correlation function \cite{LP_prd}. 

We are currently working on using the LP as a standard ruler for future galaxy surveys such as Euclid, DESI and WFIRST. These datasets will provide percent-precision measurements of the galaxy correlation function at BAO scales, increasing the relative impact of the potential bias due to non-linear effects. We have shown here that the LP provides a simple clean standard ruler that avoids many of the limitations of standard BAO methods. This may also be of interest for BAO measurements in 21-cm intensity maps that suffer from poor angular resolution, which broadens the correlation function on BAO scales but may leave the LP unaffected \cite{2017MNRAS.466.2736V}.

Several lines of investigation are currently in progress to test the validity of the LP below $1\%$, explore statistical and systematics errors, optimize the LP extraction, and explore new applications to estimate the growth of structure.

\textit{Acknowledgments.}
We are deeply grateful to Antonio Cuesta for providing us the correlation function data, as well as for his patience in answering our many questions. SA and GDS are partially supported by a Department of Energy grant DE-SC0009946 to the particle astrophysics theory group at CWRU. The research leading to these results has received funding from the European Research Council under the European Community Seventh Framework Programme (FP7/2007-2013 Grant Agreement no. 279954) ERC-StG "EDECS". IZ is supported by National Science Foundation grant AST-1612085.

Funding for SDSS-III has been provided by the Alfred P. Sloan Foundation, the Participating Institutions, the National Science Foundation, and the U.S. Department of Energy Office of Science. The SDSS-III web site is http://www.sdss3.org/.

\bibliography{MyBib}

\begin{thebibliography}{22}
\expandafter\ifx\csname natexlab\endcsname\relax\def\natexlab#1{#1}\fi
\expandafter\ifx\csname bibnamefont\endcsname\relax
  \def\bibnamefont#1{#1}\fi
\expandafter\ifx\csname bibfnamefont\endcsname\relax
  \def\bibfnamefont#1{#1}\fi
\expandafter\ifx\csname citenamefont\endcsname\relax
  \def\citenamefont#1{#1}\fi
\expandafter\ifx\csname url\endcsname\relax
  \def\url#1{\texttt{#1}}\fi
\expandafter\ifx\csname urlprefix\endcsname\relax\def\urlprefix{URL }\fi
\providecommand{\bibinfo}[2]{#2}
\providecommand{\eprint}[2][]{\url{#2}}

\bibitem[{\citenamefont{Bassett and Hlozek}(2009)}]{Bassett:2009mm}
\bibinfo{author}{\bibfnamefont{B.~A.} \bibnamefont{Bassett}} \bibnamefont{and}
  \bibinfo{author}{\bibfnamefont{R.}~\bibnamefont{Hlozek}}
  (\bibinfo{year}{2009}), \eprint{0910.5224}.

\bibitem[{\citenamefont{Eisenstein et~al.}(2005)}]{2005ApJ...633..560E}
\bibinfo{author}{\bibfnamefont{D.~J.} \bibnamefont{Eisenstein}}
  \bibnamefont{et~al.}, \bibinfo{journal}{\apj} \textbf{\bibinfo{volume}{633}},
  \bibinfo{pages}{560} (\bibinfo{year}{2005}), \eprint{astro-ph/0501171}.

\bibitem[{\citenamefont{Crocce and Scoccimarro}(2008)}]{Crocce:2007dt}
\bibinfo{author}{\bibfnamefont{M.}~\bibnamefont{Crocce}} \bibnamefont{and}
  \bibinfo{author}{\bibfnamefont{R.}~\bibnamefont{Scoccimarro}},
  \bibinfo{journal}{Phys. Rev.} \textbf{\bibinfo{volume}{D77}},
  \bibinfo{pages}{023533} (\bibinfo{year}{2008}), \eprint{0704.2783}.

\bibitem[{\citenamefont{{Smith} et~al.}(2008)\citenamefont{{Smith},
  {Scoccimarro}, and {Sheth}}}]{2008PhRvD..77d3525S}
\bibinfo{author}{\bibfnamefont{R.~E.} \bibnamefont{{Smith}}},
  \bibinfo{author}{\bibfnamefont{R.}~\bibnamefont{{Scoccimarro}}},
  \bibnamefont{and} \bibinfo{author}{\bibfnamefont{R.~K.}
  \bibnamefont{{Sheth}}}, \bibinfo{journal}{Phys. Rev. D}
  \textbf{\bibinfo{volume}{77}}, \bibinfo{eid}{043525} (\bibinfo{year}{2008}),
  \eprint{astro-ph/0703620}.

\bibitem[{\citenamefont{{Rasera} et~al.}(2014)\citenamefont{{Rasera},
  {Corasaniti}, {Alimi}, {Bouillot}, {Reverdy}, and
  {Balm{\`e}s}}}]{2014MNRAS.440.1420R}
\bibinfo{author}{\bibfnamefont{Y.}~\bibnamefont{{Rasera}}},
  \bibinfo{author}{\bibfnamefont{P.-S.} \bibnamefont{{Corasaniti}}},
  \bibinfo{author}{\bibfnamefont{J.-M.} \bibnamefont{{Alimi}}},
  \bibinfo{author}{\bibfnamefont{V.}~\bibnamefont{{Bouillot}}},
  \bibinfo{author}{\bibfnamefont{V.}~\bibnamefont{{Reverdy}}},
  \bibnamefont{and}
  \bibinfo{author}{\bibfnamefont{I.}~\bibnamefont{{Balm{\`e}s}}},
  \bibinfo{journal}{\mnras} \textbf{\bibinfo{volume}{440}},
  \bibinfo{pages}{1420} (\bibinfo{year}{2014}), \eprint{1311.5662}.

\bibitem[{\citenamefont{Anselmi and Pietroni}(2012)}]{Anselmi:2012cn}
\bibinfo{author}{\bibfnamefont{S.}~\bibnamefont{Anselmi}} \bibnamefont{and}
  \bibinfo{author}{\bibfnamefont{M.}~\bibnamefont{Pietroni}},
  \bibinfo{journal}{JCAP} \textbf{\bibinfo{volume}{1212}}, \bibinfo{pages}{013}
  (\bibinfo{year}{2012}), \eprint{1205.2235}.

\bibitem[{\citenamefont{Angulo et~al.}(2008)\citenamefont{Angulo, Baugh, Frenk,
  and Lacey}}]{Angulo:2007fw}
\bibinfo{author}{\bibfnamefont{R.}~\bibnamefont{Angulo}},
  \bibinfo{author}{\bibfnamefont{C.}~\bibnamefont{Baugh}},
  \bibinfo{author}{\bibfnamefont{C.}~\bibnamefont{Frenk}}, \bibnamefont{and}
  \bibinfo{author}{\bibfnamefont{C.}~\bibnamefont{Lacey}},
  \bibinfo{journal}{Mon. Not. R. Astron. Soc.} \textbf{\bibinfo{volume}{383}},
  \bibinfo{pages}{755} (\bibinfo{year}{2008}), \eprint{astro-ph/0702543}.

\bibitem[{\citenamefont{{Anselmi} et~al.}(2016)\citenamefont{{Anselmi},
  {Starkman}, and {Sheth}}}]{2016MNRAS.455.2474A}
\bibinfo{author}{\bibfnamefont{S.}~\bibnamefont{{Anselmi}}},
  \bibinfo{author}{\bibfnamefont{G.~D.} \bibnamefont{{Starkman}}},
  \bibnamefont{and} \bibinfo{author}{\bibfnamefont{R.~K.}
  \bibnamefont{{Sheth}}}, \bibinfo{journal}{\mnras}
  \textbf{\bibinfo{volume}{455}}, \bibinfo{pages}{2474} (\bibinfo{year}{2016}),
  \eprint{1508.01170}.

\bibitem[{\citenamefont{{Seo} et~al.}(2008)\citenamefont{{Seo}, {Siegel},
  {Eisenstein}, and {White}}}]{2008ApJ...686...13S}
\bibinfo{author}{\bibfnamefont{H.-J.} \bibnamefont{{Seo}}},
  \bibinfo{author}{\bibfnamefont{E.~R.} \bibnamefont{{Siegel}}},
  \bibinfo{author}{\bibfnamefont{D.~J.} \bibnamefont{{Eisenstein}}},
  \bibnamefont{and} \bibinfo{author}{\bibfnamefont{M.}~\bibnamefont{{White}}},
  \bibinfo{journal}{\apj} \textbf{\bibinfo{volume}{686}}, \bibinfo{pages}{13}
  (\bibinfo{year}{2008}), \eprint{0805.0117}.

\bibitem[{\citenamefont{{Eisenstein} et~al.}(2007)\citenamefont{{Eisenstein},
  {Seo}, {Sirko}, and {Spergel}}}]{2007ApJ...664..675E}
\bibinfo{author}{\bibfnamefont{D.~J.} \bibnamefont{{Eisenstein}}},
  \bibinfo{author}{\bibfnamefont{H.-J.} \bibnamefont{{Seo}}},
  \bibinfo{author}{\bibfnamefont{E.}~\bibnamefont{{Sirko}}}, \bibnamefont{and}
  \bibinfo{author}{\bibfnamefont{D.~N.} \bibnamefont{{Spergel}}},
  \bibinfo{journal}{\apj} \textbf{\bibinfo{volume}{664}}, \bibinfo{pages}{675}
  (\bibinfo{year}{2007}), \eprint{astro-ph/0604362}.

\bibitem[{\citenamefont{{S{\'a}nchez} et~al.}(2009)\citenamefont{{S{\'a}nchez},
  {Crocce}, {Cabr{\'e}}, {Baugh}, and {Gazta{\~n}aga}}}]{2009MNRAS.400.1643S}
\bibinfo{author}{\bibfnamefont{A.~G.} \bibnamefont{{S{\'a}nchez}}},
  \bibinfo{author}{\bibfnamefont{M.}~\bibnamefont{{Crocce}}},
  \bibinfo{author}{\bibfnamefont{A.}~\bibnamefont{{Cabr{\'e}}}},
  \bibinfo{author}{\bibfnamefont{C.~M.} \bibnamefont{{Baugh}}},
  \bibnamefont{and}
  \bibinfo{author}{\bibfnamefont{E.}~\bibnamefont{{Gazta{\~n}aga}}},
  \bibinfo{journal}{\mnras} \textbf{\bibinfo{volume}{400}},
  \bibinfo{pages}{1643} (\bibinfo{year}{2009}), \eprint{0901.2570}.

\bibitem[{\citenamefont{{Padmanabhan} et~al.}(2012)\citenamefont{{Padmanabhan},
  {Xu}, {Eisenstein}, {Scalzo}, {Cuesta}, {Mehta}, and
  {Kazin}}}]{2012MNRAS.427.2132P}
\bibinfo{author}{\bibfnamefont{N.}~\bibnamefont{{Padmanabhan}}},
  \bibinfo{author}{\bibfnamefont{X.}~\bibnamefont{{Xu}}},
  \bibinfo{author}{\bibfnamefont{D.~J.} \bibnamefont{{Eisenstein}}},
  \bibinfo{author}{\bibfnamefont{R.}~\bibnamefont{{Scalzo}}},
  \bibinfo{author}{\bibfnamefont{A.~J.} \bibnamefont{{Cuesta}}},
  \bibinfo{author}{\bibfnamefont{K.~T.} \bibnamefont{{Mehta}}},
  \bibnamefont{and} \bibinfo{author}{\bibfnamefont{E.}~\bibnamefont{{Kazin}}},
  \bibinfo{journal}{\mnras} \textbf{\bibinfo{volume}{427}},
  \bibinfo{pages}{2132} (\bibinfo{year}{2012}), \eprint{1202.0090}.

\bibitem[{\citenamefont{{Smith} et~al.}(2007)\citenamefont{{Smith},
  {Scoccimarro}, and {Sheth}}}]{2007PhRvD..75f3512S}
\bibinfo{author}{\bibfnamefont{R.~E.} \bibnamefont{{Smith}}},
  \bibinfo{author}{\bibfnamefont{R.}~\bibnamefont{{Scoccimarro}}},
  \bibnamefont{and} \bibinfo{author}{\bibfnamefont{R.~K.}
  \bibnamefont{{Sheth}}}, \bibinfo{journal}{\prd}
  \textbf{\bibinfo{volume}{75}}, \bibinfo{eid}{063512} (\bibinfo{year}{2007}),
  \eprint{astro-ph/0609547}.

\bibitem[{\citenamefont{{Xu} et~al.}(2012)\citenamefont{{Xu}, {Padmanabhan},
  {Eisenstein}, {Mehta}, and {Cuesta}}}]{2012MNRAS.427.2146X}
\bibinfo{author}{\bibfnamefont{X.}~\bibnamefont{{Xu}}},
  \bibinfo{author}{\bibfnamefont{N.}~\bibnamefont{{Padmanabhan}}},
  \bibinfo{author}{\bibfnamefont{D.~J.} \bibnamefont{{Eisenstein}}},
  \bibinfo{author}{\bibfnamefont{K.~T.} \bibnamefont{{Mehta}}},
  \bibnamefont{and} \bibinfo{author}{\bibfnamefont{A.~J.}
  \bibnamefont{{Cuesta}}}, \bibinfo{journal}{\mnras}
  \textbf{\bibinfo{volume}{427}}, \bibinfo{pages}{2146} (\bibinfo{year}{2012}),
  \eprint{1202.0091}.

\bibitem[{\citenamefont{Sanchez et~al.}(2012)\citenamefont{Sanchez, Scoccola,
  Ross, Percival, Manera et~al.}}]{Sanchez:2012sg}
\bibinfo{author}{\bibfnamefont{A.~G.} \bibnamefont{Sanchez}},
  \bibinfo{author}{\bibfnamefont{C.}~\bibnamefont{Scoccola}},
  \bibinfo{author}{\bibfnamefont{A.}~\bibnamefont{Ross}},
  \bibinfo{author}{\bibfnamefont{W.}~\bibnamefont{Percival}},
  \bibinfo{author}{\bibfnamefont{M.}~\bibnamefont{Manera}},
  \bibnamefont{et~al.}, \bibinfo{journal}{Mon.Not.Roy.Astron.Soc.}
  \textbf{\bibinfo{volume}{425}}, \bibinfo{pages}{415} (\bibinfo{year}{2012}),
  \eprint{1203.6616}.

\bibitem[{\citenamefont{{Xu} et~al.}(2013)\citenamefont{{Xu}, {Cuesta},
  {Padmanabhan}, {Eisenstein}, and {McBride}}}]{2013MNRAS.431.2834X}
\bibinfo{author}{\bibfnamefont{X.}~\bibnamefont{{Xu}}},
  \bibinfo{author}{\bibfnamefont{A.~J.} \bibnamefont{{Cuesta}}},
  \bibinfo{author}{\bibfnamefont{N.}~\bibnamefont{{Padmanabhan}}},
  \bibinfo{author}{\bibfnamefont{D.~J.} \bibnamefont{{Eisenstein}}},
  \bibnamefont{and} \bibinfo{author}{\bibfnamefont{C.~K.}
  \bibnamefont{{McBride}}}, \bibinfo{journal}{\mnras}
  \textbf{\bibinfo{volume}{431}}, \bibinfo{pages}{2834} (\bibinfo{year}{2013}),
  \eprint{1206.6732}.

\bibitem[{\citenamefont{{Cuesta} et~al.}(2016)\citenamefont{{Cuesta},
  {Vargas-Maga{\~n}a}, {Beutler}, {Bolton}, {Brownstein}, {Eisenstein},
  {Gil-Mar{\'{\i}}n}, {Ho}, {McBride}, {Maraston}
  et~al.}}]{2016MNRAS.457.1770C}
\bibinfo{author}{\bibfnamefont{A.~J.} \bibnamefont{{Cuesta}}},
  \bibinfo{author}{\bibfnamefont{M.}~\bibnamefont{{Vargas-Maga{\~n}a}}},
  \bibinfo{author}{\bibfnamefont{F.}~\bibnamefont{{Beutler}}},
  \bibinfo{author}{\bibfnamefont{A.~S.} \bibnamefont{{Bolton}}},
  \bibinfo{author}{\bibfnamefont{J.~R.} \bibnamefont{{Brownstein}}},
  \bibinfo{author}{\bibfnamefont{D.~J.} \bibnamefont{{Eisenstein}}},
  \bibinfo{author}{\bibfnamefont{H.}~\bibnamefont{{Gil-Mar{\'{\i}}n}}},
  \bibinfo{author}{\bibfnamefont{S.}~\bibnamefont{{Ho}}},
  \bibinfo{author}{\bibfnamefont{C.~K.} \bibnamefont{{McBride}}},
  \bibinfo{author}{\bibfnamefont{C.}~\bibnamefont{{Maraston}}},
  \bibnamefont{et~al.}, \bibinfo{journal}{\mnras}
  \textbf{\bibinfo{volume}{457}}, \bibinfo{pages}{1770} (\bibinfo{year}{2016}),
  \eprint{1509.06371}.

\bibitem[{\citenamefont{{Alam} et~al.}(2017)}]{2017MNRAS.470.2617A}
\bibinfo{author}{\bibfnamefont{S.}~\bibnamefont{{Alam}}} \bibnamefont{et~al.},
  \bibinfo{journal}{\mnras} \textbf{\bibinfo{volume}{470}},
  \bibinfo{pages}{2617} (\bibinfo{year}{2017}), \eprint{1607.03155}.

\bibitem[{\citenamefont{{Anselmi} et~al.}(2017)\citenamefont{{Anselmi},
  {Corasaniti}, {Starkman}, {Sheth}, and {Zehavi}}}]{LP_prd}
\bibinfo{author}{\bibfnamefont{S.}~\bibnamefont{{Anselmi}}},
  \bibinfo{author}{\bibfnamefont{P.-S.} \bibnamefont{{Corasaniti}}},
  \bibinfo{author}{\bibfnamefont{G.~D.} \bibnamefont{{Starkman}}},
  \bibinfo{author}{\bibfnamefont{R.~K.} \bibnamefont{{Sheth}}},
  \bibnamefont{and} \bibinfo{author}{\bibfnamefont{I.}~\bibnamefont{{Zehavi}}}
  (\bibinfo{year}{2017}), \eprint{To Appear}.

\bibitem[{\citenamefont{{Planck Collaboration}
  et~al.}(2016)\citenamefont{{Planck Collaboration}, {Ade}
  et~al.}}]{2016A&A...594A..13P}
\bibinfo{author}{\bibnamefont{{Planck Collaboration}}},
  \bibinfo{author}{\bibfnamefont{P.~A.~R.} \bibnamefont{{Ade}}},
  \bibnamefont{et~al.}, \bibinfo{journal}{\aap} \textbf{\bibinfo{volume}{594}},
  \bibinfo{eid}{A13} (\bibinfo{year}{2016}), \eprint{1502.01589}.

\bibitem[{\citenamefont{{Lewis} et~al.}(2000)\citenamefont{{Lewis},
  {Challinor}, and {Lasenby}}}]{2000ApJ...538..473L}
\bibinfo{author}{\bibfnamefont{A.}~\bibnamefont{{Lewis}}},
  \bibinfo{author}{\bibfnamefont{A.}~\bibnamefont{{Challinor}}},
  \bibnamefont{and}
  \bibinfo{author}{\bibfnamefont{A.}~\bibnamefont{{Lasenby}}},
  \bibinfo{journal}{\apj} \textbf{\bibinfo{volume}{538}}, \bibinfo{pages}{473}
  (\bibinfo{year}{2000}), \eprint{astro-ph/9911177}.

\bibitem[{\citenamefont{{Villaescusa-Navarro}
  et~al.}(2017)\citenamefont{{Villaescusa-Navarro}, {Alonso}, and
  {Viel}}}]{2017MNRAS.466.2736V}
\bibinfo{author}{\bibfnamefont{F.}~\bibnamefont{{Villaescusa-Navarro}}},
  \bibinfo{author}{\bibfnamefont{D.}~\bibnamefont{{Alonso}}}, \bibnamefont{and}
  \bibinfo{author}{\bibfnamefont{M.}~\bibnamefont{{Viel}}},
  \bibinfo{journal}{\mnras} \textbf{\bibinfo{volume}{466}},
  \bibinfo{pages}{2736} (\bibinfo{year}{2017}), \eprint{1609.00019}.

\end{thebibliography}
 
\end{document}